
\texsis
\draft
\WorldScientific

\def\memu{{M(e\mu)}}

\def\journal#1&#2(#3)#4{{\unskip,~\sl #1\unskip~\bf\ignorespaces #2\unskip~\rm
(19#3) #4}}
\def\jour#1&#2(#3)#4{{\unskip ~\sl #1\unskip~\bf\ignorespaces #2\unskip~\rm
(19#3) #4}}

{15 August 1993 \hfill MSUTH 93/11}
\title{On Top Quark Polarization and Proposed Techniques for
Measuring~$m_t$~at~Hadron~Colliders}
\author
Glenn~A.~Ladinsky
{\it Michigan State University, Department of Physics and Astronomy}
{\it East Lansing, MI 48824}
\endauthor

\abstract
{\tenpoint
Measuring the top quark mass in its semileptonic decay mode, $t\to
be^+\nu$ ($b\to\mu^-+X$),
through the mass distribution of the $e^+$--$\mu^-$ pair has
been shown to provide a good measurement of the top quark mass with an
uncertainty of $1.6\%$ for $m_t=150,250\,$GeV at the SSC.  Here it is
demonstrated that an uncertainty in the
polarization of the top quark can add an additional uncertainty of
about $3\%$ in the mass.}
\endabstract
\footnote{}{\ninerm\baselineskip=11pt Talk presented at the Workshop on
Physics at Current Accelerators and the Supercollider held at the
Argonne National Laboratory, Argonne, Illinois, 2-5 June 1993.}
%
%
\referencelist
\reference{sdctech}
Solenoidal Detector Collaboration, {\it Technical Design of a Detector
to be operated at the Superconducting Super Collider}, 1 April 1992,
preprint SDC-92-201
\endreference
\reference{kly}
C.~Kao, G.~A. Ladinsky and C.--P. Yuan, unpublished
\endreference
\reference{tpol}
G.L.~Kane, G.A.~Ladinsky and C.--P.~Yuan
\journal Phys.~Rev. &D45 (92) 124
\endreference
\reference{*doug}
D.O.~Carlson and C.--P.~Yuan \journal Phys.~Lett.&B306 (93) 386
\endreference
\reference{gold}
R.H.~Dalitz and G.R.~Goldstein \journal Phys.Rev.&D45 (92) 1531
\endreference
\reference{slacfrag}
C.~Peterson, D.~Schlatter, I.~Schmitt and P.M.~Zerwas
\journal Phys.~Rev.&D27 (83) 105
\endreference
\reference{aleph}
ALEPH Collaboration (D.~DeCamp, et al.) \journal Phys.~Lett.&B244 (90) 551
\endreference
\reference{peskin}
C.R.~Schmidt and M.E.~Peskin \journal Phys.~Rev.~Lett.&69 (92) 410
\endreference
\endreferencelist
%
%

Among the techniques proposed for measuring the mass of the top quark ($m_t$)
at the Superconducting Supercollider (SSC),
is the determination of $m_t$ in the semileptonic decay
mode $t\to be^+\nu$, where after hadronization and decay, the bottom
quark ($b$) yields a $\mu^-$ from which the mass of the $e^+$--$\mu^-$ system,
$M(e\mu)$, provides the information needed to resolve $m_t$.
For $m_t=150,\ 250\,$GeV the precision obtained in measuring the top quark
mass is $1.6\%$.\cite{sdctech}

In work the author performed in collaboration with C.~Kao and
C.--P.~Yuan,\cite{kly} it was mentioned how polarization can have an
effect on distributions used to make precision measurements.  In
particular, it was noted in this reference how the mass distribution
of the $b-e^+$ system ($M(eb)$) in $t\to be^+\nu$ varied with the
polarization of the top quark.  This presentation goes one step
further, taking the fragmentation of the $b$ quark into account, and
finds that if polarization is not properly applied, the
uncertainty in the $m_t$ measurement grows.


In \Ref{sdctech} the ISAJET Monte Carlo was used to study the
production of top quarks at the SSC
($pp$ collisions at $40\,$TeV with a luminosity of $10\hbox{fb}^{-1}$/yr).
Kinematic constraints were applied to approximate the
detector geometry, reduce background
and increase the observable's sensitivity to $m_t$.
Among these cuts, all visible
particles were restricted in rapidity ($\eta$) to $|\eta|<2.5$, the
transverse momentum of the $\mu^-$ was restricted
to be greater than $20\,$GeV and the transverse momentum of the
$e^+$ was chosen to be greater
than $40\,$GeV.  The combined transverse momentum of the $e^+$ and
$\mu^-$ was required to be greater than $100\,$GeV for $m_t=150\,$GeV
and the difference in azimuthal angle ($\phi$) between those two
leptons was restricted to be less than $80^o$ to suppress
background which could appear
from mistakenly using the muon appearing from the charm decay associated
with the other top quark in the event.
An isolation cut on the positron was invoked, where
the sum of the transverse energy within a cone of
$\Delta R=\sqrt{(\Delta\phi)^2+(\Delta\eta)^2}$ about the $e^+$ had to be less
than $4\,$GeV, while a non--isolation cut on the muon required
the sum of the transverse energy within a cone of $\Delta R<0.2$
about the $\mu^-$
to be greater than $20\,$GeV.
It was from studying the production of unpolarized top quarks in this
manner that the an error estimate on the top quark mass of $\delta
m_t=1.6\%$ was obtained.


About $10^8$ $t\bar{t}$ pairs are expected at the SSC with $gg\to t\bar{t}$
being the main production mode, but
top quark polarization can come from various sources, {\it e.g.},
electroweak production, electroweak radiative corrections,
undetected emission of particles with parity violating
interactions, or new physics.  Even at tree level in the standard model
the top quark is completely polarized when produced via
$gW^+\to\bar{b}t$.\cite{tpol}
Looking at the amplitude squared for the decay of the
top quark in terms the momentum of the $\nu$, $b$, and $e^+$
($p_\nu$, $p_b$ and $p_e$, respectively),\cite{kly}\cite{gold}
$$
|M|^2=
{64G_F^2m_W^4\over (p_W^2-m_W^2)^2+m_W^2\Gamma_W^2}
(p_\nu\cdot p_b)[(p_e\cdot p_t)-m_t(p_e\cdot s_t)],
\EQN{ampsq}
$$
where the $W$ boson width is $\Gamma_W$ and its mass is $m_W$,
it is apparent that the $\memu$ distribution is going to vary according to the
top quark spin, $s_t$.  If no kinematic constraints were applied,
the spin dependence would vanish upon integration over the phase space,
but since experiments apply cuts on the kinematics, a spin effect may remain.

\figure{fboth}
\epsfxsize=6in
\centerline{\epsfbox{fig.eps}}
\caption{This plot displays the distribution
$d\sigma\over dM(e\mu)$~vs.~$M(e\mu)$ for the separate top quark helicities
in the tree level production of $t\bar t$ pairs at the SSC using
$m_t=150\,$GeV.}
\endfigure

A parton level Monte Carlo with the Peterson fragmentation
parameterization\cite{slacfrag} (using $\epsilon_b=0.006$ from \Ref{aleph})
was used to produce \Fig{fboth}, but without
the isolation cuts used in \Ref{sdctech}.  Before the cuts, the event
rates for producing left--handed and right--handed top quarks in
$t\bar{t}$ pairs were about the same.  In \Fig{fboth} there is a
preference for keeping events with top quarks that carry
right--handed helicity.  This is because a right--handed top quark
preferentially decays with the positron moving along the top quark
direction of motion while the opposite is true for a left--handed top
quark.\cite{peskin}  The boost the top quark carries therefore
produces more energetic positrons such that the large $p_T$ cut on the
positron eliminates more events
with left--handed top quarks than with right--handed top quarks.

\midtable{tone}
\caption{\tenpoint  Though for left--handed and right--handed top quark
helicities the width remains about the same
for the gaussian fits to the $M(e\mu)$ distribution,
a shift in the peak appears.  The relative height values are normailzed
against the left--handed top quark of $m_t=150\,$GeV.}
\singlespaced
\ruledtable
\multispan4\hfil GAUSSIAN FITS TO $M(e\mu)$ \hfil\CR
$m_t=150\,$GeV \CR
Top Quark Helicity  \dbl  Mean     | Width            |  Relative Height   \CR
Left--Handed        \dbl  $50\,$GeV   | $17\,$GeV    |    $1.00$    \cr
Right--Handed       \dbl  $46\,$GeV   | $16\,$GeV    |    $1.85$     \CR
$m_t=180\,$GeV \CR
Top Quark Helicity  \dbl  Mean        | Width         |  Relative Height   \CR
Left--Handed        \dbl  $66\,$GeV   | $21\,$GeV     | $0.68$      \cr
Right--Handed       \dbl  $60\,$GeV   | $22\,$GeV     | $0.95$
\endruledtable
\endtable

The two most important features of the $\memu$ distribution in
determining $m_t$ are the kinematic maximum and the location of the
peak.  The statistics decrease away from the maximum and the
fragmentation makes the high end of the $\memu$ distribution less
sharp than the high end of the $M(eb)$ distribution.\cite{kly}
Fitting the curves with a Gaussian distribution as done in
\Ref{sdctech}, \Tbl{tone} shows that a shift in the peak occurs on the
order of $6-10\%$ when comparing completely left--handed top decays with
completely right--handed top decays.  The width does not vary much with
polarization, but it is also less sensitive to $m_t$.
So, using the location of the peak and interpolating
between the values in \Tbl{tone}, the uncertainty between left--handed and
right--handed polarizations roughly provides a $3\%$ uncertainty in $m_t$.

As the top quark mass grows, the kinematic cuts become less selective
between left--handed and right--handed helicities (as seen in the
relative heights of \Tbl{tone})
and the $\memu$ distributions for the two helicities
increase in similarity
in the decline they
exhibit in the high mass region.  \Tbl{tone}, however,
still indicates a significant shift in the peak for $m_t=180\,$GeV.

In conclusion, it is found that
the $\memu$ distribution remains a good tool for the precision
measurement of the top quark mass at the SSC, particularly where the
dominant production mode can be described through a parity conserving
process, such as $gg\to t\bar t$ in QCD.  Provided the
polarization introduced by electroweak corrections remains small,
there will be only a negligible effect to consider when studying $m_t$.
If a high level of polarization appears, whether it be through new
physics or through the specific examination of standard model processes
such as $Wg\to \bar{b}t$, one should make certain that any comparisons
between theory and experiment
correctly account for polarization in the decay of the
top quark, or else there may be a significant loss in precision.

I thank C.~Kao and C.--P.~Yuan for a pleasant collaboration.
This work was funded in part by TNRLC grant \#RGFY9240.

%
%
\nosechead{References}

\ListReferences
\vfill\eject
\end